# Predicting Survival of Tongue Cancer Patients by Machine Learning Models

Angelos Vasilopoulos [a] and Nan Miles Xi [a, *]

[a] *Department of Mathematics and Statistics, Loyola University Chicago, Chicago, IL 60660, USA*

[*] Correspondence: Nan Miles Xi (mxi1@luc.edu)

## Abstract

Tongue cancer is a common oral cavity malignancy that originates in the mouth and throat. Much effort has been invested in improving its diagnosis, treatment, and management. Surgical removal, chemotherapy, and radiation therapy remain the major treatment for tongue cancer. The patient's survival determines the treatment effect. Previous studies have identified certain survival and risk factors based on descriptive statistics, ignoring the complex, nonlinear relationship among clinical and demographic variables. In this study, we utilize five cutting-edge machine learning models and clinical data to predict the survival of tongue cancer patients after treatment. Five-fold cross-validation, bootstrap analysis, and permutation feature importance are applied to estimate and interpret model performance. The prognostic factors identified by our method are consistent with previous clinical studies. Our method is accurate, interpretable, and thus useable as additional evidence in tongue cancer treatment and management.

# Introduction

Tongue cancer is one of the most frequent head and neck malignancies. According to the American Cancer Society, tongue cancer is diagnosed in approximately 20,000 patients and causes more than 2,700 deaths annually in the United States [1]. The average age of diagnosis is around 63, and 20% of cases occur in patients younger than 55 [2]. The overall rate of new cases has risen in the last 20 years due to smoking, drinking alcohol, and human papillomavirus infection, the three major risk factors [3]. The clinical community groups tongue cancer into two types based on its location: oral cancer, beginning in the front two-thirds of the tongue, and oropharyngeal cancer, beginning at the back third of the tongue [3]. Tongue cancer typically originates in the squamous cells that line the tongue's surface, and the types of cells affected may determine prognosis and treatment [4]. Tongue cancer treatment primarily involves surgical removal, chemotherapy, and radiation therapy [5]. The 5-year relative survival rate after treatment was 68.8% between 2012 and 2018 [6].

Given tongue cancer's wide occurrence, accurate survival prediction is crucial for treatment design and management. In recent years, machine learning has been successfully applied in related medical fields, including cancer therapy [7], drug development [8,9], and precision medicine [10]. Machine learning is potentially effective in predicting survival from patient data and identifying important factors in treatment. However, the literature has few machine learning models of tongue cancer survival. Although some studies have identified certain survival and risk factors, the conclusions are usually based on descriptive statistics and linear models, ignoring the complex, nonlinear relationship among clinical and demographic variables [2,11,12].

In this paper, we utilize a comprehensive machine learning framework to predict tongue cancer survival after treatment. The analysis is performed on a real clinical dataset containing information on 1712 patients receiving curative tongue cancer surgery. We train five cutting-edge machine learning models on this dataset and provide unbiased prediction performance by five-fold cross-validation. We further quantify the uncertainty of model performance by bootstrap analysis. The important prognostic factors in model prediction are identified by permutation feature importance. Overall, the proposed machine learning models show high accuracy in predicting patient survival. The prediction is consistent in terms of point estimation and uncertainty measurement. The identified prognostic factors echo previous findings in clinical studies. Our method is accurate, interpretable, and thus useable as additional evidence in tongue cancer treatment and management.

## Dataset and preprocessing

In this study, we analyze a dataset collected from 1712 tongue cancer curative surgery recipients at Chang Gung Memorial Hospitals, Taiwan, from 2004 to 2013 [2]. Among all patients, 1280 survive after surgery (74.77%), and 432 fail to do so (25.23%). Survival information is recorded at follow-up time for each patient. The median follow-up time of all patients is 2.88 years. We treat survival as positive and non-survival as negative in our analysis. The original dataset contains 12 variables of patients' medical records and demographic information. The area of operation and occurrence of operation are the same for all subjects and therefore do not contribute to classification. We exclude these variables from analysis along with follow-up time, which is not known prior to patient survival. Eight variables are used to predict patient survival, including tumor stage, T stage, N stage, tumor grade, radiation therapy, chemotherapy, gender, and age. The summary statistics of all variables are shown in Table 1.

Tumor stage is assigned according to the American Joint Committee on Cancer's (AJCC) TNM classification of malignant tumors [13]: in stage 0, tumors have not grown or spread; in stage 1, tumors are small and have not spread; in stage 2, tumors have grown but not spread; in stage 3, tumors are larger and have spread to surrounding tissue or lymph nodes of the immune system; in stages 4A, 4B, and 4C, tumors are larger and have spread to at least one other body organ (secondary or metastatic cancer). T stage is the size of a tumor, labeled by numbers one (small) to four (large). N stage indicates whether cancer has metastasized to lymph nodes, labeled by numbers zero (no metastasis to lymph nodes) to three (metastasis to multiple lymph nodes). Tumor grade is associated with the rate of cancer metastasis to other organs, represented by grades one (cancer cells resemble normal cells and are not proliferating) to three (cancer cells look abnormal and spread aggressively). Radiation therapy is a binary variable, indicating the administration of ionizing radiation to control or kill malignant tumor cells. Chemotherapy is also a binary variable, indicating a regimen of one or multiple anti-cancer drugs.

# Methods

In machine learning terminology, the prediction of patient survival is a binary classification task. We utilize k-nearest neighbors (kNN) [14], random forest [15], extreme gradient boosting (XGBoost) [16], logistic regression with a $L_2$ penalty (logistic LASSO regression) [14], and an ensemble of these four models to tackle this task using the dataset described in the last section. kNN classifies patient survival according to Euclidian proximity to neighbors. Random forest classifies patient survival according to the most frequent outcome of a set of decision trees. Like random forest, XGBoost assigns patient survival according to the outcomes of multiple decision trees; however, trees are grown sequentially, and each tree learns from the ones preceding it. Logistic LASSO regression regularizes the coefficients of logistic regression by an $L_2$ penalty, shrinking some to zero and thereby accomplishing feature selection. An ensemble model is a combination of the aforementioned four models, i.e., the survival probabilities output by four models is averaged before conversion to patient survival labels. We refer to the four non-ensemble models as individual models moving forward.

We evaluate the model performance by five measurements: accuracy, precision, recall, true negative rate (TNR), and area under the precision-recall curve (AUPRC). Accuracy is the ratio of true predictions to the total number of patients; precision is the ratio of true positive predictions to all positive predictions; recall is the ratio of true positive predictions to all positive patients; TNR is the ratio of true negative predictions to all negative patients; AUPRC measures the overall capacity of a binary predictive model and adjusts for class imbalance in the data.

We calculate the five measurements of each model by five-fold cross-validation. First, patients are split into five groups, or folds. Then, five different combinations of four folds are used as training sets in each of five successive cross-validation iterations, with the remaining fold in each iteration as a test set for performance assessment. Finally, we average performance measurements over five iterations. We also conduct a grid search to finetune hyperparameters in the four individual models. All performance measurements are calculated using the hyperparameter combinations with the highest AUPRC under five-fold cross-validation. The hyperparameters we finetune for each model and their optimal values are summarized in Table 2.

## Results

**Overall prediction performance.** Table 3 summarizes the accuracy, AUPRC, precision, recall, and TNR of the five models proposed in the last section. All measurements are calculated by five-fold cross-validation using the optimal hyperparameters in Table 2. Among four individual models, XGBoost achieves the highest accuracy (0.7664), AUPRC (0.8802), precision (0.7855), and TNR (0.2335), indicating a solid overall capacity of differentiating between positive and negative patients. Random forest outperforms others in recall (0.9752), showing its strength in identifying surviving patients. Combining the four individual models as an ensemble model improves AUPRC over the best-performing dividual model. It also provides close-to-top performance in terms of accuracy, precision, and recall. Overall, the machine learning models show mixed performance in predicting patient survival. There is no single model dominating others in all five measurements. The leading performance of XGBoost and random forest demonstrate the strong nonlinear relationship between patient survival and other variables, which are largely ignored in previous studies. The ensemble model balances different measurements by utilizing the strengths of individual models.

**Flexibility between positive and negative predictions.** The overall performance in Table 3 shows that all models have high recalls (above 0.9) but low TNRs (below 0.3), indicating an imbalance in predicting positive and negative patients. The reason is that all models, by default, use 0.5 as the probability cutoff for assigning binary outcome labels. A patient with a positive probability greater than 0.5 is predicted as positive (survival); otherwise, negative (not survival). The 0.5 cutoff forces the models to predict more positive patients than negative patients, resulting in low TNRs. To examine the models' flexibility between positive and negative predictions, we adjust the probability cutoff from 0.5 to 0.9, with a step size of 0.1, and then calculate the corresponding precisions, recalls, and TNRs, respectively (Table 4). Under larger cutoffs, models predict fewer positive patients and more negative patients, resulting in lower recalls but higher TNRs. For example, Table 4 shows that when the cutoff is changed from 0.5 to 0.9, the TNR of XGBoost increases from 0.2335 to 0.9910, and its recall decreases from 0.9463 to 0.0399. Model users can choose appropriate cutoffs based on their interest in positive or negative predictions. Another benefit of larger cutoffs is that they improve the accuracy of predicted positives, i.e., higher precision, through more cautious positive prediction. As the cutoff increases from 0.5 to 0.9, Table 4 shows that the precision of XGBoost increases from 0.7855 to 0.9398.

**Prediction uncertainty measurement.** The five measurements in Table 3 are point estimations of model performance. We further estimate the uncertainty of model prediction by bootstrapping. Specifically, we repeat five-fold cross-validation 1000 times, with training data resampled with replacement in every iteration. Each resampling is used to train one of the previous five models before assessing accuracy, AUPRC, precision, recall, and TNR on the test set in cross-validation. All models use their optimal hyperparameters from Table 2. We determine the model uncertainty from the resulting empirical distribution of five-fold measurements. Figure 1 presents visual comparisons of measurement distributions for each model. Table 5 includes empirical 95% confidence intervals and means of performance measurements from bootstrapping. The five models show a slightly different asymptotical performance ranking compared with their point estimation in Table 2. Logistic LASSO regression has the highest mean accuracy, precision, and TNR. The ensemble model outperforms other models on mean AUPRC and recall. The performance differences among the five models are moderate in accuracy and AUPRC, the two overall measurements. However, the gaps are more significant in precision, recall, and TNR, indicating diverse model behavior in distinguishing positive and negative patients. We also observe less performance variation in logistic LASSO regression and ensemble model, an expected result given the stable model structure of logistic LASSO regression and the diverse model components of the ensemble model.

**Feature importance analysis.** We utilize the permutation feature importance [15] to measure the contribution of each variable to survival prediction. Permutation feature importance is the decrease of AUPRC when the model predicts a test set with one variable permuted. Because permutation breaks the relationship between variables and patient survival, a subsequent decrease in AUPRC indicates model dependency on that variable for prediction. For each variable, we average its permutation feature importance across five models. We then divide those averages by the largest importance among all variables to obtain the normalized permutation feature importance. Figure 2 represents variable ranking from most important to least important in terms of normalized permutation feature importance. Tumor grade contributes the most to the prediction of patient survival, followed by N stage, T stage, chemotherapy, and radiation therapy. These variables are consistent with previous findings in clinical and modeling studies. For example, histological grading and the TNM staging system (i.e., tumor grade, N stage, and T stage) are well-established prognostic factors in oral cancer diagnosis and treatment [17,18]. Numerous studies also suggest the importance of adjuvant therapy (i.e., chemotherapy and radiation therapy), especially for patients in advanced stages [17,19]. On the other hand, age and gender are among the least important variables in the model prediction, making less than 10% of the contribution of the top variable, tumor grade. These two variables are also found to have less impact on patient survival in clinical studies [20,21].

## Conclusions

In this study, we utilize machine learning models to predict the survival of tongue cancer patients after receiving curative surgery. Our models are built on a clinical dataset with 1712 patients. We use five measurements to provide a comprehensive evaluation of model performance. Although no individual model outperforms others in all measurements, the nonlinear models, i.e., XGBoost and random forest, exhibit better overall accuracy. The linear predictive model, logistic LASSO regression, provides more stable prediction in bootstrap analysis. We also find that the ensemble model improves accuracy and stability by incorporating the strength of individual models. By adjusting the probability cutoff, our models offer flexibility in predicting positive and negative patients. Our feature importance analysis identifies key variables in predicting patient survival, consistent with previous findings in clinical and modeling studies. Overall, the machine learning models show satisfactory prediction performance. The average accuracy and AUPRC of the five models are 0.7588 and 0.8785, respectively. In practice, AUPRC greater than 0.8 indicates excellent discrimination between binary outcomes, especially given the imbalanced dataset in our study [22].

Several topics are worth exploring in future studies. First, all 1712 patients in the current dataset are from Taiwan. Collecting larger datasets from a more diverse population will increase the generality of machine learning models for new patients. Second, additional variables about patients' lifestyles and physical features, including smoking habits, body mass index, and occupational history, could contribute to creating superior models. Third, emerging genomic technology, e.g., single-cell RNA-sequencing (scRNA-seq), can be applied to reveal the transcriptomes of tongue cancer patients and identify molecular biomarkers[23–27]. Finally, our machine learning framework can evaluate the quality of clinical data in the survival diagnosis of tongue cancer. A high-quality dataset contains clinical information for models to accurately predict patient survival. The model prediction accuracy can serve as a proxy for the data quality of different datasets.

## Data and Code Availability

The data used in this study is available at Zenodo repository:

https://zenodo.org/record/7450476#.Y532FHaZMuJ

The source code that implemented the result in this study is available at GitHub repository:

https://github.com/angvasilop/tongue_cancer

# Tables

Table 1. Summary statistics of all variables used in this study.

| Variable | Value | Percentage |
|---|---|---|
| Age (years) | | |
| Mean ± SD | 51.83 ± 11.29 | |
| Range | 21 – 92 | |
| Gender | | |
| Male | 1504 | 87.85% |
| Female | 208 | 12.15% |
| Stage | | |
| 1 | 611 | 35.69% |
| 2 | 406 | 23.71% |
| 3 | 233 | 13.61% |
| 4 | 1 | 0.06% |
| 4A | 453 | 26.46% |
| 4B | 6 | 0.35% |
| 4C | 2 | 0.12% |
| T stage | | |
| 1 | 657 | 38.38% |
| 2 | 629 | 36.74% |
| 3 | 156 | 9.11% |
| 4 | 270 | 15.77% |
| N stage | | |
| 0 | 1202 | 70.21% |
| 1 | 180 | 10.51% |
| 2 | 327 | 19.10% |
| 3 | 3 | 0.18% |
| Grade | | |
| 1 | 516 | 30.14% |
| 2 | 1033 | 60.34% |
| 3 | 163 | 9.52% |
| Radiation therapy | | |
| Yes | 612 | 35.75% |
| No | 1100 | 64.25% |
| Chemotherapy | | |
| Yes | 469 | 27.39% |
| No | 1243 | 72.61% |
| Survival | | |
| Yes | 1280 | 74.77% |
| No | 423 | 25.23% |

**Table 2. Optimal hyperparameters for the four individual models.** Each value is determined by five-fold cross-validation with AUPRC as the optimization criterion.

| Model | Hyperparameter | Optimal Values | Description |
|---|---|---|---|
| kNN | k | 85 | Number of neighbors |
| Random forest | ntree | 100 | Number of trees |
|  | mtry | 1 | Number of variables sampled |
|  | nodesize | 3 | Minimum observations in a terminal node |
| XGBoost | nrounds | 7 | Maximum number of iterations |
|  | max.depth | 2 | Tree depth |
|  | $\eta$ | 0.5 | Learning rate |
| Logistic LASSO regression | $\lambda$ | 0.02 | Shrinkage coefficient |

**Table 3. Five measurements of model prediction performance.** Each measurement is calculated by five-fold cross-validation using optimal hyperparameters. The highest values among the five models are underscored.

| Model | Accuracy | AUPRC | Precision | Recall | TNR |
|---|---|---|---|---|---|
| kNN | 0.7523 | 0.8726 | 0.7658 | 0.9652 | 0.1236 |
| Random forest | 0.7593 | 0.8791 | 0.7675 | <u>0.9752</u> | 0.1237 |
| XGBoost | <u>0.7664</u> | 0.8802 | <u>0.7855</u> | 0.9463 | <u>0.2335</u> |
| Logistic LASSO regression | 0.7553 | 0.8752 | 0.7820 | 0.9347 | 0.2257 |
| Ensemble | 0.7605 | <u>0.8855</u> | 0.7719 | 0.9658 | 0.1538 |

**Table 4. Three measurements of model performance under different cutoffs.** The probability cutoff of positive patients is adjusted from 0.5 to 0.9. Then corresponding precisions, recalls, and TNRs are calculated for each model.

| Model | Measurement | Cutoff | | | | |
| --- | --- | --- | --- | --- | --- | --- |
| | | 0.5 | 0.6 | 0.7 | 0.8 | 0.9 |
| kNN | Precision | 0.7658 | 0.7962 | 0.8355 | 0.8657 | 0.9013 |
| | Recall | 0.9652 | 0.8918 | 0.7853 | 0.6254 | 0.3603 |
| | TNR | 0.1236 | 0.3210 | 0.5406 | 0.7126 | 0.8836 |
| Random forest | Precision | 0.7675 | 0.7896 | 0.8094 | 0.8336 | 0.8562 |
| | Recall | 0.9752 | 0.9394 | 0.8948 | 0.8400 | 0.7548 |
| | TNR | 0.1237 | 0.2570 | 0.3725 | 0.5004 | 0.6208 |
| XGBoost | Precision | 0.7855 | 0.8277 | 0.8491 | 0.8779 | 0.9398 |
| | Recall | 0.9463 | 0.8570 | 0.7557 | 0.6650 | 0.0399 |
| | TNR | 0.2335 | 0.4677 | 0.6005 | 0.7229 | 0.9910 |
| Logistic LASSO regression | Precision | 0.7820 | 0.8212 | 0.8549 | 0.8720 | 0.0000 |
| | Recall | 0.9347 | 0.8711 | 0.7596 | 0.6790 | 0.0000 |
| | TNR | 0.2257 | 0.4359 | 0.6165 | 0.7023 | 1.0000 |
| Ensemble | Precision | 0.7719 | 0.8113 | 0.8339 | 0.8638 | 0.9062 |
| | Recall | 0.9658 | 0.8954 | 0.8227 | 0.7158 | 0.3380 |
| | TNR | 0.1538 | 0.3808 | 0.5122 | 0.6624 | 0.8963 |

**Table 5. Summary statistics of model performance calculated by bootstrapping.** The empirical 95% confidence intervals and means of five measurements are calculated for each model. The highest values among the five models are underscored.

| Model | Measurement | Accuracy | AUPRC | Precision | Recall | TNR |
|---|---|---|---|---|---|---|
| kNN | 95% CI | (0.7430, 0.7623) | (0.8533, 0.8752) | (0.7599, 0.7814) | (0.9338, 0.9776) | (0.0886, 0.2224) |
|  | Mean | 0.7531 | 0.8649 | 0.7698 | 0.9570 | 0.1504 |
| Random forest | CI | (0.7465, 0.7658) | (0.8605, 0.8737) | (0.7593, 0.7855) | (0.9320, 0.9818) | (0.0815, 0.2403) |
|  | Mean | 0.7558 | 0.8674 | 0.7713 | 0.9593 | 0.1549 |
| XGBoost | CI | (0.7477, 0.7693) | (0.8544, 0.8771) | (0.7773, 0.8011) | (0.9024, 0.9496) | (0.1974, 0.3312) |
|  | Mean | 0.7588 | 0.8665 | 0.7886 | 0.9270 | 0.2623 |
| Logistic LASSO regression | CI | (0.7535, 0.7699) | (0.8568, 0.8620) | (0.7851, 0.8042) | (0.9045, 0.9360) | (0.2438, 0.3440) |
|  | Mean | <u>0.7622</u> | 0.8595 | <u>0.7951</u> | 0.9197 | <u>0.2973</u> |
| Ensemble | CI | (0.7500, 0.7652) | (0.8711, 0.8834) | (0.7639, 0.7752) | (0.9569, 0.9757) | (0.1130, 0.1732) |
|  | Mean | 0.7583 | <u>0.8776</u> | 0.7697 | <u>0.9668</u> | 0.1431 |

# Figures

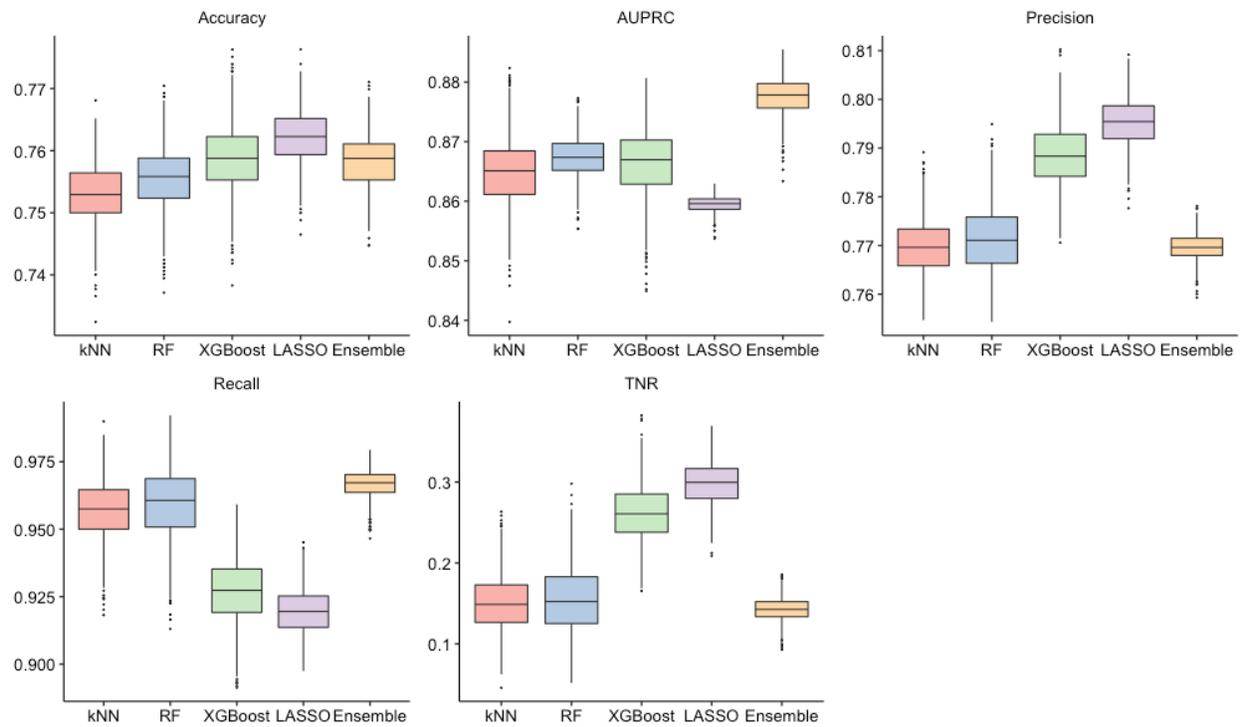

**Figure 1. The empirical distributions of model performance computed by bootstrapping.** Five performance measurements of the five models are obtained from 1000 bootstrap iterations.

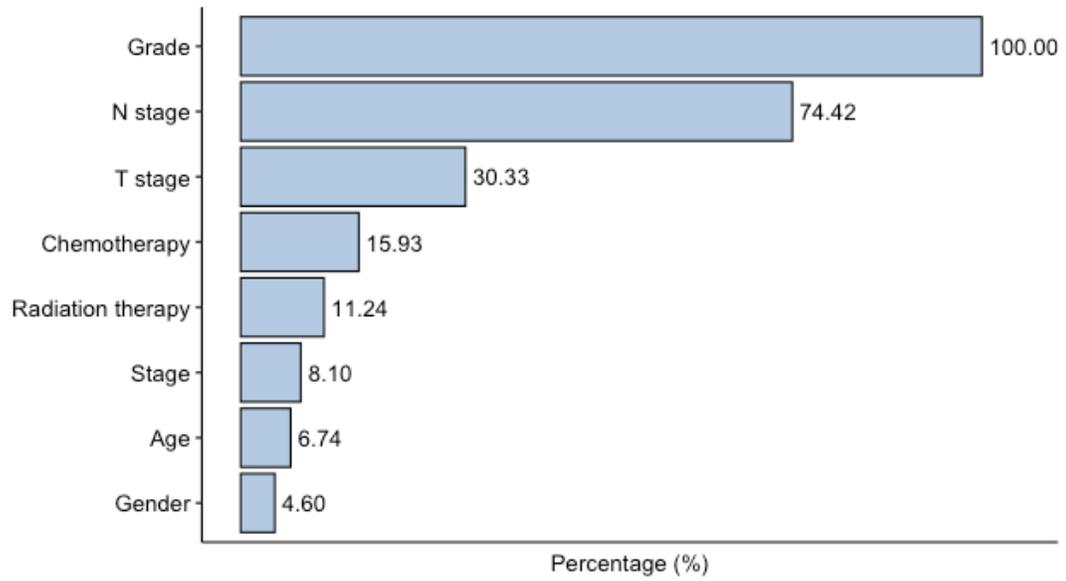

**Figure 2. The normalized permutation feature importance**. Variables are sorted from highest to lowest normalized importance.